\documentclass[10pt,conference]{IEEEtran}
\usepackage[utf8]{inputenc}

\usepackage{amssymb}
\usepackage{amsthm}
\usepackage{float}
\usepackage{graphics}
\usepackage{graphicx}
\usepackage{epstopdf}
\usepackage{url}
\usepackage{verbatim}
\usepackage{xcolor}
\usepackage[draft]{todonotes}   
\usepackage[font=small]{caption} 
\newcommand{\subparagraph}{}
\usepackage{titlesec}
\titlespacing{\section}{0pt}{0.60\baselineskip}{0.60\baselineskip} 
\titlespacing{\subsection}{5pt}{0.30\baselineskip}{0.30\baselineskip}
\begin{document}
\title{A Centralized SDN Architecture for the 5G Cellular Network}
\author{
\IEEEauthorblockN{Akshatha Nayak M., Pranav Jha, Abhay Karandikar}
\IEEEauthorblockA{Department of Electrical Engineering, Indian Institute of Technology Bombay \\
Email: {$\lbrace$akshatha, pranavjha, karandi$\rbrace$}@ee.iitb.ac.in}
}
\maketitle
\bibliographystyle{IEEEtran} 
\begin{abstract}
 In order to meet the increasing demands of high data rate and low latency cellular broadband applications, plans are underway to roll out the Fifth Generation (5G) cellular wireless system by the year 2020. This paper proposes a novel method for adapting the Third Generation Partnership Project (3GPP)'s 5G architecture to the principles of Software Defined Networking (SDN). We propose to have centralized network functions in the 5G network core to control the network, end-to-end. This is achieved by relocating the control functionality present in the 5G Radio Access Network (RAN) to the network core, resulting in the conversion of the base station known as the gNB into a pure data plane node. This brings about a significant reduction in signaling costs between the RAN and the core network. It also results in improved system performance. The merits of our proposal have been illustrated by evaluating the Key Performance Indicators (KPIs) of the 5G network, such as network attach (registration) time and handover time. We have also demonstrated improvements in attach time and system throughput due to the use of centralized algorithms for mobility management with the help of ns-3 simulations.
\end{abstract}

\section{Introduction}
The number of cellular broadband subscriptions has been growing steadily since the last decade. The Ericsson mobility report predicts that subscriptions will increase to 8.3 billion by the year 2022 from 4.4 billion in 2016~\cite{ericsson}. There has also been a considerable increase in the per capita data consumption due to the popularity of data-intensive applications like video streaming, augmented reality, etc.. At the same time, applications with diverse latency and devices with varied power and throughput requirements are becoming increasingly common within the network \cite{5gamericas}. All these developments are ushering in newer challenges for control and management of existing cellular networks. Some of these aspects are being addressed by the Fifth Generation (5G) cellular wireless system, the standardization for which is currently underway in Third Generation Partnership Project (3GPP)~\cite{5gspec}.
  
\par Software Defined Networking (SDN) and Network Function Virtualization (NFV), have been proposed as two of the key enablers for 5G cellular wireless networks~\cite{5gspec}. SDN \cite{sdnrfc} is a networking paradigm that introduces an abstraction between the control and data planes. The control plane comprises of protocols that control and manage network devices. Network devices that carry data traffic constitute the data plane. SDN provides standardized interfaces between the two planes. Standardization of device interfaces simplifies network management by enabling the use of uniform policy based rules and eliminating the need for vendor specific configurations. 

\par The system architecture for the 5G cellular network \cite{5gspec} as defined by 3GPP, marks a departure from the fourth generation Long Term Evolution (LTE) architecture by restructuring network elements as network functions. Network functions interact with each other over well-defined interfaces. These can be classified as control or data plane functions, with a few exceptions, e.g., the New Radio (NR) based 5G Radio Access Network (RAN) function~\cite{5granspec}. The RAN function as embodied in NR NodeB (gNB) possesses both control and data plane functionalities.

\par In this paper, we present a new architecture for the 5G cellular network which extends the SDN paradigm to the RAN function. The proposed architecture centralizes the control function for the complete system and places it in the network core. This is achieved by moving the RAN control functions, i.e., Radio Resource Control (RRC) protocol layer and the Radio Resource Management (RRM) functionalities from the gNB to the core network. As a result, the gNB is transformed into a node containing only data plane functionality, managed through a standard interface from the centralized control function located in the core network. The restructuring of the gNB results in a significant reduction in signaling between the RAN and the network core. The improvements obtained over the 3GPP defined 5G architecture have been illustrated with the help of callflow comparisons. There is also a reduction in signaling failure scenarios e.g., reduction in handover failures and improvement in the overall system performance due to the centralization of network control. We present the performance analysis for both the network architectures. The results of the analysis have been corroborated with the help of simulations using ns-3~\cite{ns3}, a network simulation software.

\par The rest of the paper is organized as follows:
Section II summarizes the related work that has been carried out in this area. Sections III and IV describe 3GPP's 5G and the proposed network architectures, respectively. Section V details the performance analysis of both the network architectures. Section VI discusses simulation results. This is followed by Section VII which concludes the paper and provides areas for future work.  

\section{Related Work}
 As the architecture for the 5G cellular network is relatively new and its standardization is still in progress, a majority of the existing literature is based on the application of SDN principles to LTE networks. Also, the 5G signaling procedures defined till date~\cite{5gspec},~\cite{5granspec} and its protocol architecture bear a lot of similarity to that of the LTE network. We summarize the relevant work applying the SDN paradigm to the LTE core as well as the radio access network, while highlighting the work dealing with the application of SDN to the RAN in the following paragraph, as it bears most similarity with our work. The authors in \cite{karimzadeh2017software} propose a new architecture for a flat LTE network for achieving increased scalability, by merging the functionalities of Serving Gateway (S-GW) and Packet Data Network Gateways (P-GW). On a similar note, the authors in \cite{chourasia2015sdn} have aimed to reduce the signaling and tunneling costs in the core network by replacing the S-GW and P-GW with a single OpenFlow switch~\cite{version1}. They replace the General Packet Radio Service (GPRS) Tunneling Protocol (GTP) with OpenFlow for reducing signaling costs incurred due to tunneling. In another work \cite{basta2013virtual}, the authors have proposed several architectural solutions describing the optimal distribution of core network elements between the cloud infrastructure and the data plane for reducing operator costs and improve network performance. They achieve improvements by introducing a new network element with support for additional network functions in OpenFlow, together with a flexible placement of core network elements.
 
 \par The authors in \cite{khan2016system} define a new centralized system architecture for efficient resource management in LTE network. The proposed architecture has been shown to improve fairness, downlink throughput, and signaling reductions. The paper suggests decoupling certain key radio resource functionalities, e.g., handover functionality from the eNodeB and placing them in a centralized SDN controller. Their proposed framework keeps the control plane interface from the eNodeB towards the Mobility Management Entity (MME) intact and consequently, may not reduce the processing time for control signals. The authors in \cite{zhang2015architecture} propose a new SDN architecture for \emph{5G} networks based on the LTE system. Their architecture aims to manage the end-to-end network in a centralized manner by using separate controllers for the RAN and the core network. The RAN controller is responsible for mobility and interference management, whereas the core network controller regulates routing and policy.  
 
\par However, to the best of our knowledge, this is the first paper that proposes a centralized SDN controller architecture for the recently defined 3GPP 5G network and evaluates the benefits of the same.

\section{5G Network architecture}
\begin{figure}
\centering
\includegraphics[width=0.45\textwidth]{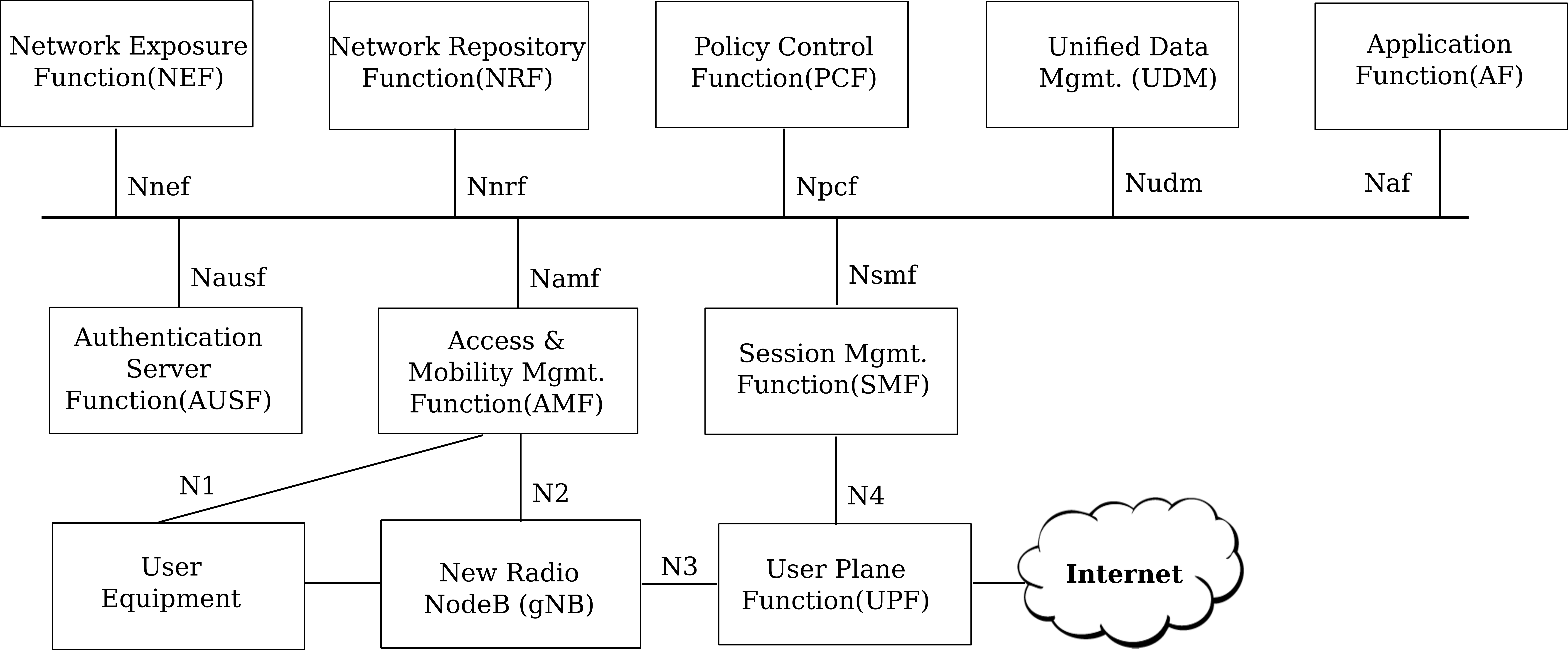}
\caption{3GPP defined 5G Network Architecture (Adapted from \cite{5gspec})}
\label{fig:5garch}
\end{figure}

 The 5G network architecture as defined by 3GPP is shown in Figure \ref{fig:5garch}. The 5G architecture is a service based architecture, wherein, the network functions interact with each other using well-defined interfaces, e.g., the RAN and Access and Mobility Function (AMF) communicate with each other using the N2 interface. The 5G cellular network has been designed to be inter-operable with the existing LTE network.

\par The 5G network consists of two parts:
\begin{enumerate}
\item Next Generation Radio Access Network (NG-RAN):
The NR based RAN for the 5G network extends between the 5G UE and the gNB. The gNB is divided into two logical nodes, viz., the gNB Central Unit (gNB-CU) and the gNB Distributed Unit (gNB-DU) \cite{5granspec}. These nodes are interconnected with one another over a data plane interface known as F1-U and the control plane interface known as F1-C. The F1 Application Protocol (F1-AP) runs over the F1-C interface. The F1-AP is used to carry messages for configuring the gNB-DU. The gNB-CU has both control and data plane functionalities and hosts the RRC, Service Data Adaptation Protocol (SDAP), Packet Data Convergence Protocol (PDCP) and Next Generation User plane (NG-U) protocols. The RRC layer along with the RRM functions is responsible for the management of connected mode mobility, security keys, enforcement of Quality of Service (QoS) on the radio interface, radio bearer control and radio admission control. The gNB-CU controls the operation of one or more gNB-DUs. The gNB-DU consists of RLC, MAC and PHY layers. The gNB-CU and gNB-DU together, appear as a unified logical entity (gNB) to the core network.
\item 5G Core (5GC): The 5GC connects the gNBs to the external data network. The 5GC consists of a multitude of network functions, some of which have been listed in Table \ref{table:5gcore} along with their salient functionalities.
\begin{table}
 \caption{5G Core Network Functions.}
 \centering
\begin{tabular}{|p{2.5cm}|p{5cm}|}
  \hline
  \textbf{Network Function}  & \textbf{Functionality} \\ 
  \hline
  Access and Mobility Management Function (AMF) & Acts as the termination point for the Non Access Stratum (NAS) signaling, mobility management \\
  \hline
  Authentication Server Function (AUSF)   &  Supports the UE authentication process\\
  \hline 
  User Plane Function (UPF) &  Serves as the anchor point for intra/inter-Radio Access Technology (RAT) mobility, packet routing, traffic reporting, handles user plane Quality of Service (QoS)\\
  \hline
   Session Management Function (SMF) & Supports the establishment, modification and release of a data session, configuration of traffic steering policies at the UPF, UE Internet Protocol(IP) address allocation and policy enforcement \\
  \hline 
\end{tabular} 
\label{table:5gcore}
\end{table} 
\end{enumerate}

\par The signaling procedures of the 5G cellular network are similar to that of LTE, since the standard considers co-deployment scenarios for the LTE Evolved Packet Core (EPC) and the 5GC. The 5G cellular network provides backward compatibility with the Evolved Terrestrial Radio Access Network (E-UTRAN), by using an enhanced LTE eNodeB, known as the next generation eNodeB (ng-eNB)~\cite{5gspec}. 

\section{Proposed Architecture}
\begin{figure}
\centering
\includegraphics[width=0.45\textwidth]{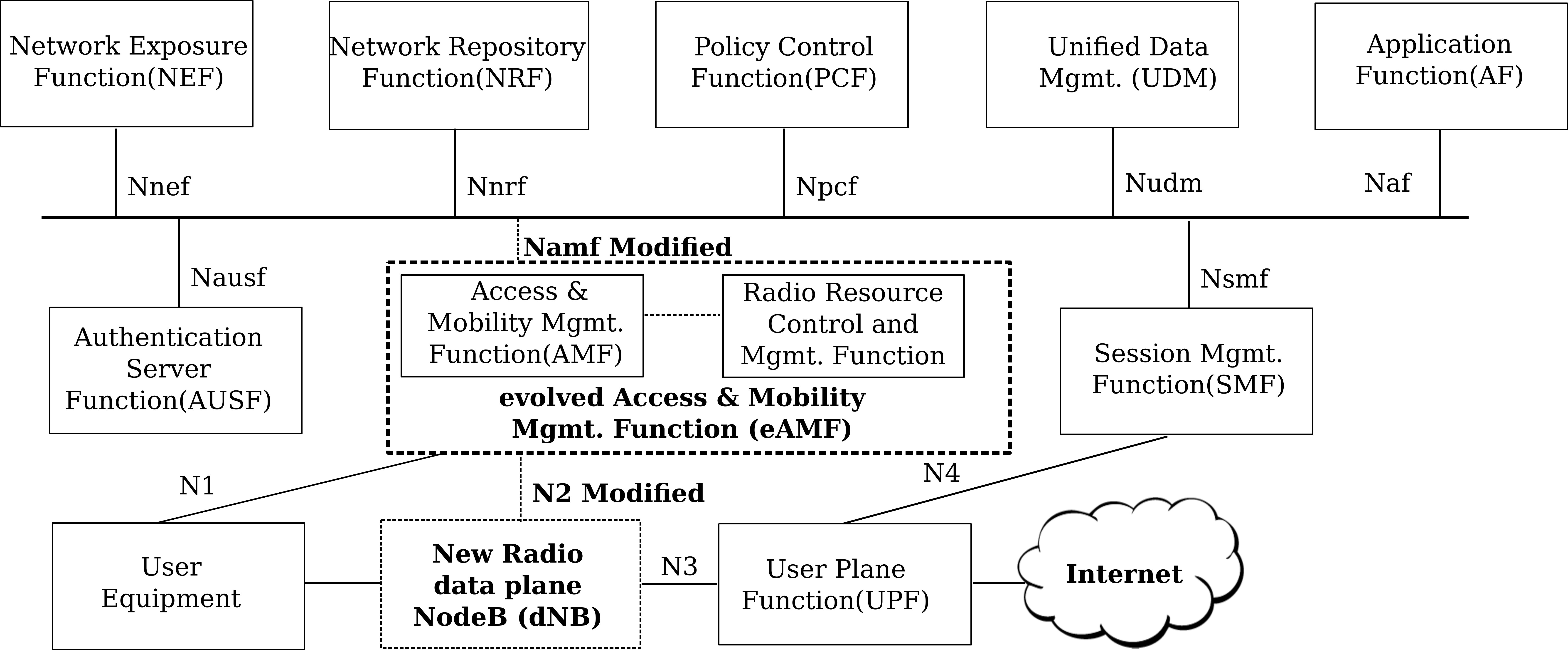}
\caption{Proposed 5G network architecture.}
\label{fig:sdn5Garch}
\end{figure}

In this section, we propose a modified architecture for the 5G network. In the proposed architecture, the control functionality of gNB, i.e., the RRC layer along with the RRM function is removed from the gNB and placed in the core network. We refer to the new gNB, devoid of control plane functionality and comprising of only data plane functionality, as the NR data plane NodeB (dNB). The RRC protocol layer and RRM functionality, together with the AMF constitute a new network function located in the core network, hereinafter referred to as the enhanced AMF (eAMF). In addition, the F1-AP which is used by gNB-CU to configure the gNB-DU in the 3GPP defined architecture, is modified and used by eAMF to control and manage the dNBs. As a result, network control gets centralized and a well-defined separation between control and data planes in the end-to-end network is achieved. Although we consider the gNB as the reference base station in our architecture, this proposal is also valid for the ng-eNB~\cite{5gspec}.

\par The placement of RRC and RRM functions in the core gives rise to several advantages:

\subsection{Reduction in signaling cost due to the elimination of NG-AP layer:}
\begin{figure}
\centering
\includegraphics[width=0.45\textwidth]{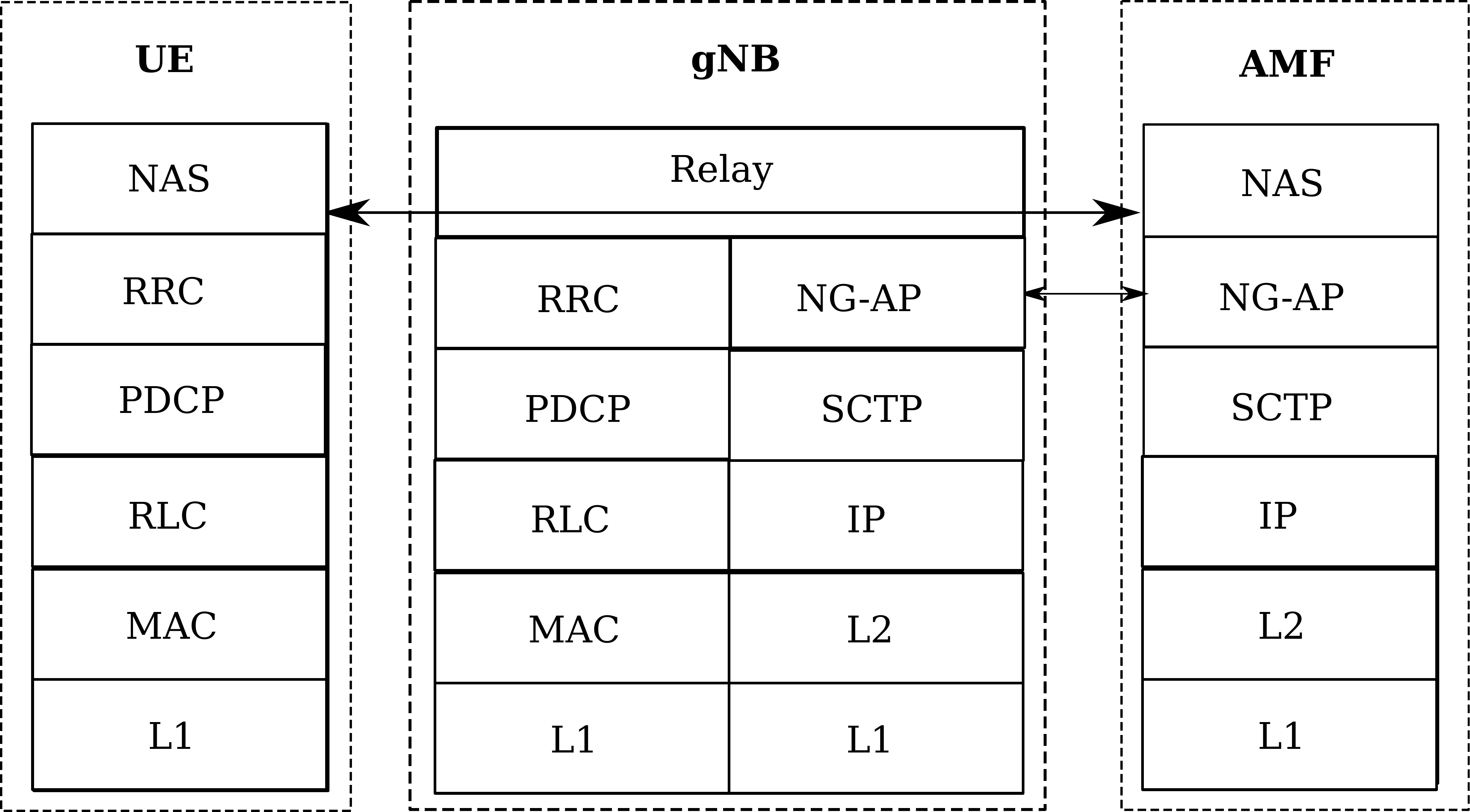}
\caption{3GPP defined 5G control plane stack.}
\label{fig:sdn5Gcontrol}
\end{figure}

Figure \ref{fig:sdn5Gcontrol} depicts the protocol stack for the 3GPP defined 5G network.
As shown in Figure \ref{fig:sdn5Gcontrol}, the gNB has a UE facing protocol stack consisting of RRC, PDCP, Radio Link Control (RLC), Medium Access Control (MAC) and Layer1 (L1) layers. The protocol stack of the gNB that interfaces with the core network consist of the Next Generation-Application Protocol (NG-AP), Stream Control Transmission Protocol (SCTP), IP, Layer 2 (L2) and L1 protocols.

\par
\begin{figure}
\centering
\includegraphics[width=0.45\textwidth]{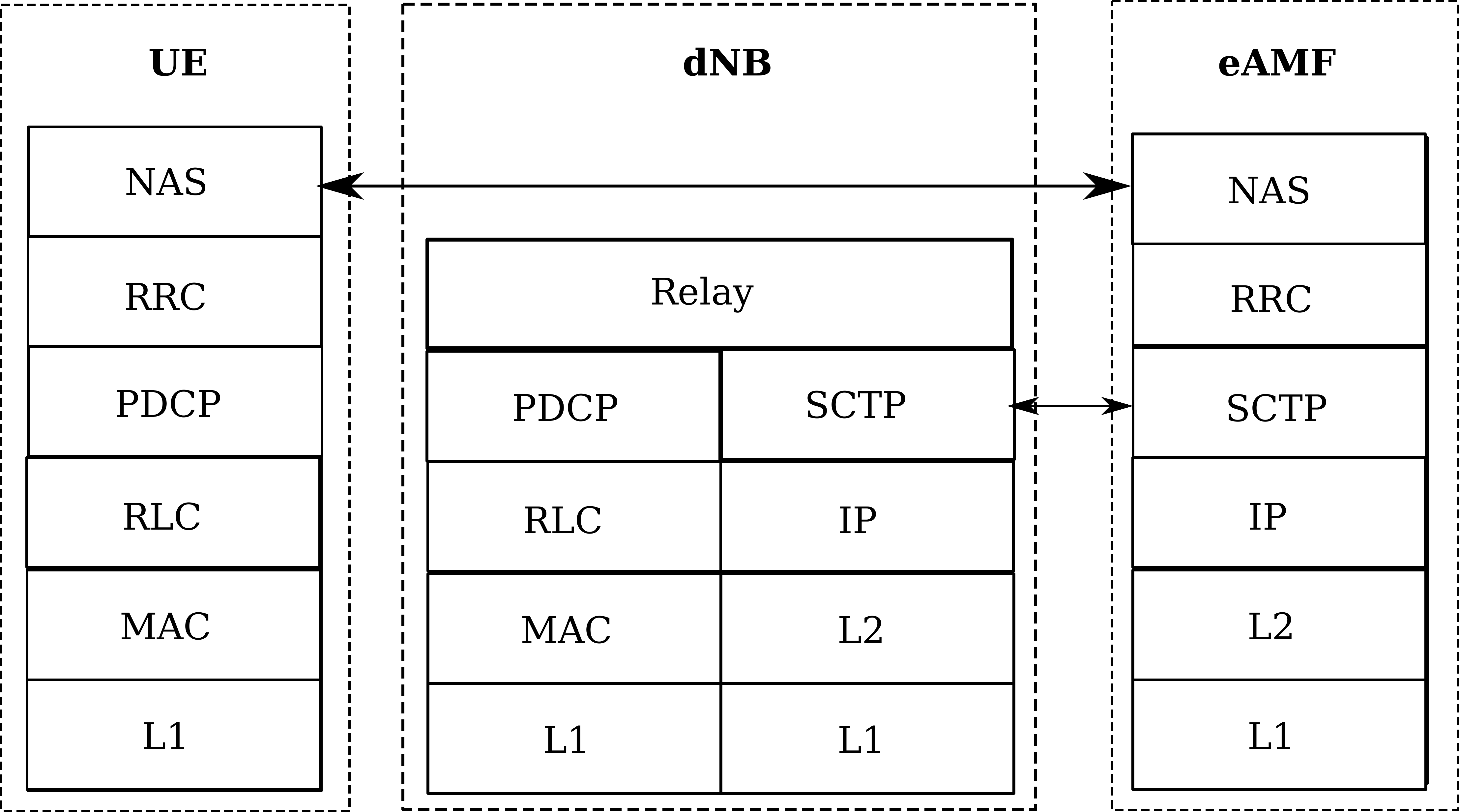}
\caption{Control plane stack for the proposed architecture.}
\label{fig:proposed5Gcontrol}
\end{figure}

 In the 3GPP defined network, the RRC layer along with the RRM function in the gNB perform radio resource allocation. Since both gNB and AMF possess control plane functionality, the NG-AP is needed for signaling exchanges between gNB and the 5GC, e.g., to carry UE specific signaling. As a result of transposing RRC along with the RRM functionality into the AMF, the control functionality is completely transferred to the core network. The NG-AP is no longer required to carry UE specific signaling between gNB and AMF and can thus be eliminated. The resultant protocol stack for the proposed architecture is shown in Figure \ref{fig:proposed5Gcontrol}. 

\par
\begin{figure}
\centering
\includegraphics[width=0.45\textwidth]{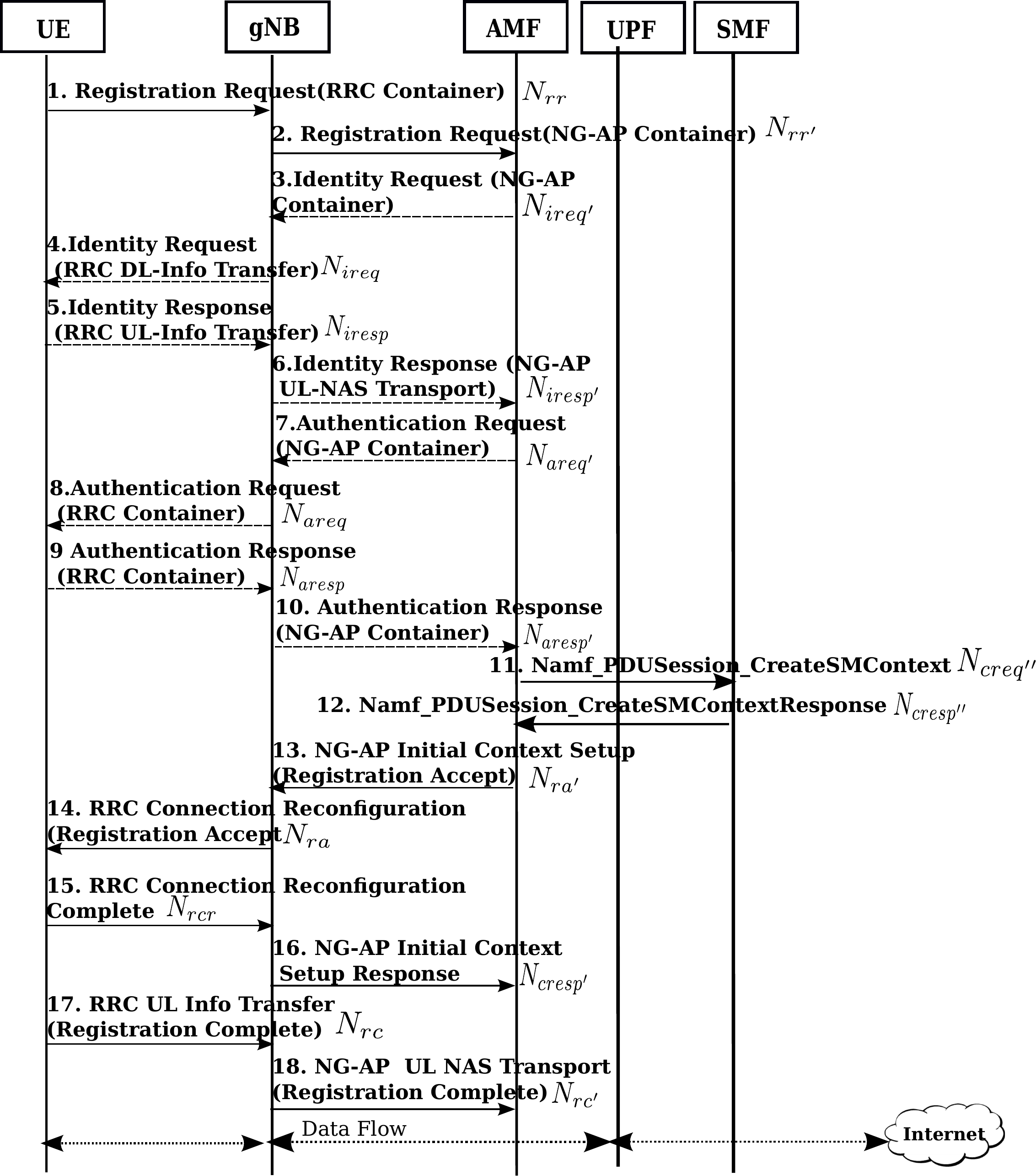}
\caption{Registration procedure for the 3GPP defined 5G architecture.}
\label{fig:5Gattach}
\end{figure}

\begin{figure}
\centering
\includegraphics[width=0.9\linewidth]{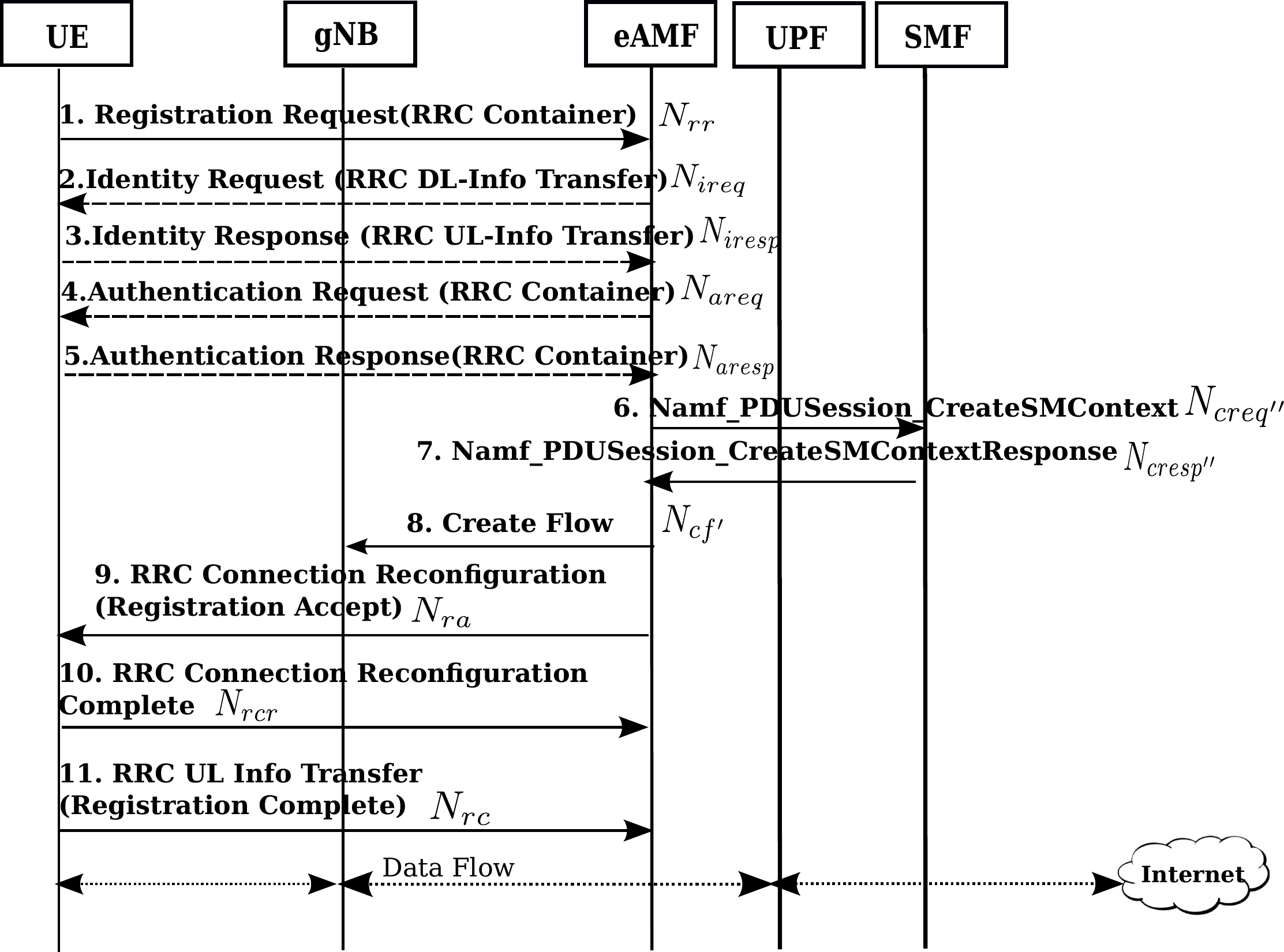}
\caption{Registration procedure for the proposed architecture.}
\label{fig:sdn5Gattach}
\end{figure}

 In order to demonstrate the advantages of the proposed architecture with respect to the 3GPP defined 5G network, we study the callflows for registration and handover for both the architectures. These call flows have been modeled using the LTE call flows as a reference. Registration is a procedure by which a UE attempts to access the cellular network for the first time. It is equivalent to the attach procedure in the LTE network. The details of the procedure for 3GPP defined 5G and the proposed networks have been illustrated in Figures \ref{fig:5Gattach} and \ref{fig:sdn5Gattach}, respectively.

\par In the 3GPP defined 5G cellular network, the registration procedure mainly involves control message exchanges between UE, gNB, and AMF. UE exchanges Non Access Stratum (NAS) messages with the AMF by encapsulating them using RRC protocol and transmitting them to the gNB. The gNB decodes the received messages and sends them further to the AMF with the help of NG-AP. As a result, every message exchanged between the UE and AMF is processed twice. In order to distinguish between messages encoded using RRC and NG-AP in Figure \ref{fig:5Gattach}, we have shown them as being encoded in RRC and NG-AP containers, respectively. Additionally, a few signaling messages are also exchanged between the gNB and AMF, to setup flow contexts on the gNB for data transfer to a particular UE. On completion of the above signaling exchanges, the data flow may be initiated in the network.

\par The callflow for the registration procedure in the proposed architecture is illustrated in Figure \ref{fig:sdn5Gattach}. All NAS and RRC messages are exchanged between the UE and eAMF, via the dNB. These messages are encoded/decoded using the RRC protocol. An additional Create Flow message is introduced. The message is used by eAMF to instruct the dNB to create a new data flow. This message is sent over a modified F1-AP to configure the dNB in accordance with the flow requirements.
 
\par By comparing both the call flows, we can infer the following: 
\begin{itemize}
 \item Due to the removal of NG-AP, some of the signaling messages resulting from encoding to NG-AP, e.g., Initial UE message etc., are eliminated from the call flow.
 \item  The number of encoding and decoding steps for identity verification, authentication etc., are reduced as the RRC messages are directly transmitted to eAMF without being processed at dNB.
 \item  As the decision making is centralized, there is no longer a need for handshake messages/acknowledgments, e.g., the NG-AP message like the Initial Context Setup Response is sent by gNB to AMF in response to the Initial Context Setup Request message in the standard 5G network. Such response messages are no longer required.
\end{itemize}

\par All of the above factors bring about a significant reduction in the signaling cost between the NG-RAN and the 5GC, thereby improving the performance of the system.

\begin{figure}
\centering
\includegraphics[width=0.9\linewidth]{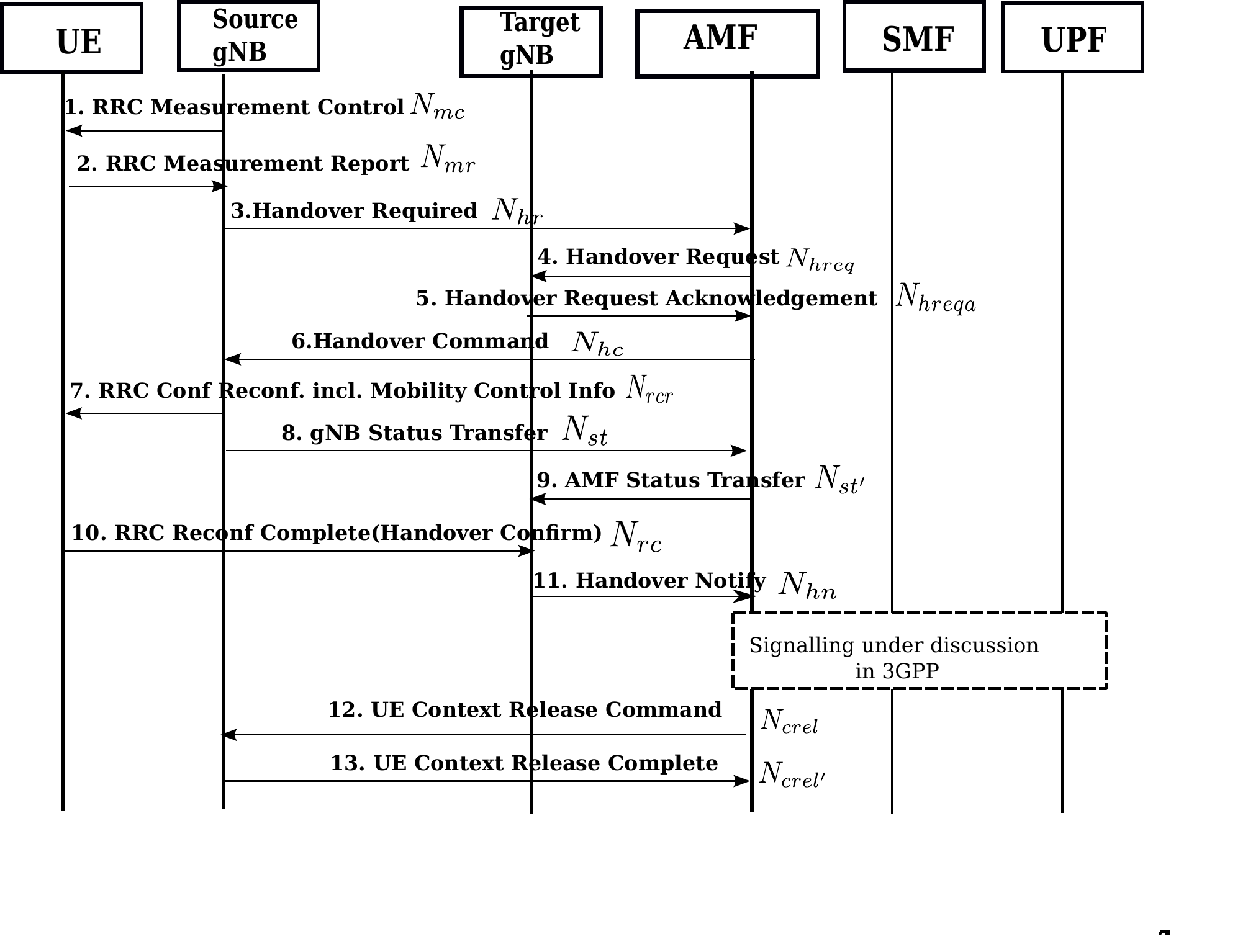}
\caption{Handover in the 3GPP defined 5G architecture.}
\label{fig:5Ghandover}
\end{figure}
 
\par
\begin{figure}
\centering
\includegraphics[width=0.45\textwidth]{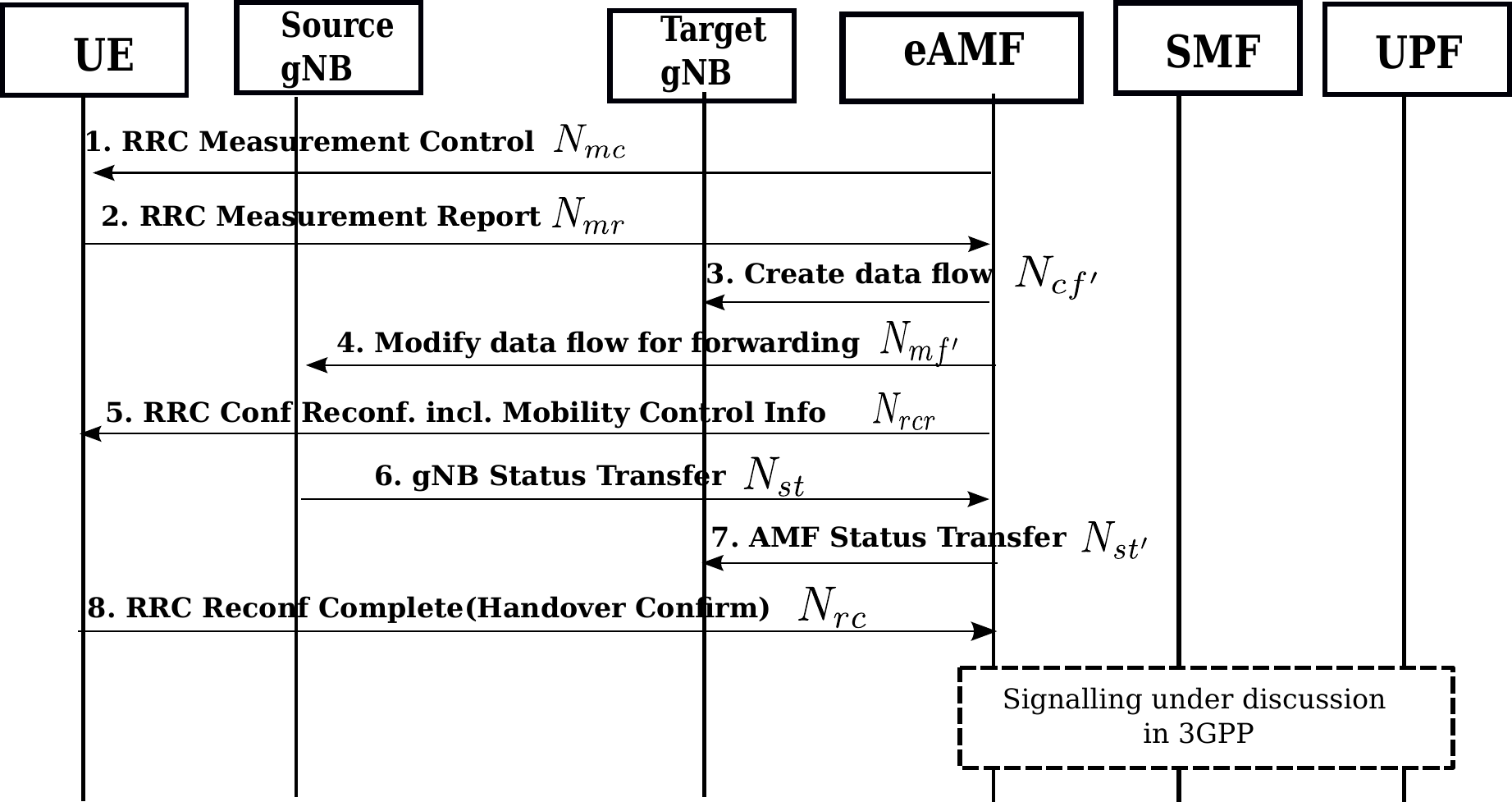}
\caption{Handover in the proposed architecture.}
\label{fig:sdn5Ghandover}
\end{figure}

 Similar conclusions can be obtained for the handover procedure, which has been shown in Figures \ref{fig:5Ghandover} and \ref{fig:sdn5Ghandover}, respectively. In the 3GPP defined 5G cellular architecture, the gNB receives RRC measurement reports from a UE and sends Handover Required message to the AMF for handover initiation, whenever required. The AMF transmits a Handover Request message to the prospective target gNB, which responds with Handover Request Acknowledgement, if it is able to admit the UE. The AMF then issues a Handover Command message to the source gNB to handover the UE to the chosen target. The source gNB sends an RRC Connection Reconfiguration message to the UE to indicate the same. The UE then sends the Handover Confirm message to the target gNB. Following this, the Handover Notify message is sent from the target gNB to the AMF. Once these steps are completed, session setup is carried out in the core network. This part of the procedure is still under study in the 3GPP working group~\cite{5gproc}. We have illustrated this step only for the sake of completion and it does not affect our analysis as the message exchanges for the session setup are within the 5G Core network and not across the core and the NG-RAN. After the completion of the session setup, the older UE context is released from the source gNB.

\par In our proposed architecture, similar to the registration procedure, all of the measurement reports are sent to the eAMF. The eAMF is responsible for handover decisions and transmits commands for data flow creation and modification to the gNBs, when necessary. The remainder of the call flow remains unchanged with the exception of the UE context release step, which is no longer required as the context is centrally stored in the eAMF. We observe that the handover signaling has been simplified due to the centralization of control.

\subsection{Reduced Mobility Failures and Faster Handover}
The centralized view of network resources also aids in making better decisions for handover management. In our proposed architecture, the eAMF maintains context for the UEs and hence has access to the overall network state, e.g., traffic load at a given gNB-DU, signal strength of various cells as observed by the UEs, UE QoS requirements, and data rates, etc.. In the 3GPP defined architecture, mobility decisions are taken at both gNBs and AMF as both UE context and the decision making abilities are distributed. Centralization of mobility management provides a network-wide view of resources and leads to reduced handover failures as well as faster handover for the UEs. 

\subsection{Load Balancing and Interference Management}
The proposed architecture can also facilitate better interference management and load balancing decisions with optimized algorithms, which bring about an increase in the overall system throughput. For example, where the traffic distribution is not uniform, the eAMF can take decisions to handover UEs from heavily loaded cells to the lightly loaded ones. This can be helped by strategies such as dNB transmit power control, cell-offset tuning, etc..

\subsection{UE Power saving:}
\par The reduction in time for network access and idle mode mobility procedures results in power savings for UEs as they can now remain in the power saving idle mode for a longer time. This is due to a reduction in the time required for the UE to switch between the idle and active states as illustrated by the registration callflows.

\subsection{Reduced system costs:}
\par Several studies advocate the placement of computationally intensive network control functions in the datacenter and time-sensitive data plane functions in the network infrastructure closer to the UE \cite{basta2013virtual}, \cite{al2015cord}. Our proposal is in alignment with this thinking and helps in reducing the costs of the gNBs, which can now be replaced with simpler devices having radio functionality. 
 
\section{Performance Analysis}
\par We present the performance analysis for the overall signaling cost reduction in this section. We use Figures \ref{fig:5Gattach}, \ref{fig:sdn5Gattach}, \ref{fig:5Ghandover} and \ref{fig:sdn5Ghandover} as the reference figures to illustrate the same. The reference figures also list out the signaling messages that are exchanged between various network elements during registration and handover, respectively. The processing overheads in a given node for both the architectures are given in Table \ref{table:process}. 

\begin{table}
\caption{ASN1 Processing Overhead}
\centering
\resizebox{0.45\textwidth}{!}{\begin{tabular}{||l |l ||} 
 \hline
  \textbf{ASN1 Messages} & \textbf{Notation}   \\ 
 \hline\hline
 RRC decode at gNB of message received from UE & $P_{gd}$   \\ 
 \hline
 RRC encode at gNB of message sent to UE & $P_{ge}$ \\
 \hline
 NG-AP encode at gNB of message sent to AMF  & $P_{ge'}$  \\
 \hline
 NG-AP decode at gNB  of message received from AMF  & $P_{gd'}$  \\
 \hline
 RRC encode at eAMF of message received from UE & $P_{ee}$   \\ 
 \hline
 RRC decode at eAMF of message sent to UE & $P_{ed}$ \\
 \hline
 F1-AP encode at eAMF of message sent to dNB & $P_{e'e}$   \\ 
 \hline
  F1-AP decode at eAMF of message sent from dNB & $P_{e'd}$   \\ 
 \hline
 F1-AP encode at dnb of message sent to eAMF & $P_{de}$ \\
 \hline
 F1-AP decode at dnb of message received from eAMF & $P_{dd}$ \\
 \hline
 \end{tabular}}
 \label{table:process}
 \end{table}
 
As the message sizes are yet to be standardized in the 5G specification, we assume that all messages have an average length of $m$ bits. We also assume that the time taken per bit for message exchange between any two nodes is $\alpha$ and the processing time for any node, mentioned in Table \ref{table:process} is equal to $\beta$. We also observe that, according to the values provided in~\cite{feasibilityspec} for $\alpha(\approx 1)$ms and $\beta(\approx 4)$ms, $\alpha < \beta$. In accordance with the above assumptions, we have calculated the time taken for signaling in the 5G architecture as well as our proposed architecture below:

\begin{enumerate}
\item Registration Signaling Time for the 3GPP 5G architecture\\
               $T_{Attach} = \alpha(N_{rr})+ 5(P_{gd}+P_{ge'})+ \alpha(Nrr') + \alpha(N_{ireq'})$
                $+ 3(P_{gd'}+P_{ge}) + \alpha(N_{ireq}+ N_{iresp} + N_{iresp'}) + \alpha(N_{areq'}+ N_{areq}+ N_{aresp} + N_{aresp'} + N_{creq''} + N_{cresp''}) + \alpha(N_{ra'} +N_{ra} +N_{rcr} + N_{cresp'} + N_{rc} + N_{rc'}) +5P_{gd}+3P_{ge}$ 
               $= 18m\alpha+ 24\beta.$
\item Registration Signaling Cost for the proposed architecture\\
               $T_{Attach'} = 2\alpha(N_{rr}+N_{ireq}+ N_{iresp}+ N_{areq} +N_{aresp}) + \beta(3P_{ee}+ 5P_{ed}) $ 
               $+\alpha(N_{creq''}+ N_{cresp''}+N_{cf'}) +2\alpha(N_{ra}+N_{rcr}+N_{rc})+P_{e'e}+P_{dd}$
               $= 19m\alpha + 10\beta.$
\item Handover signaling cost for the 3GPP 5G architecture \\
	       $T_{Handover}=\alpha(N_{mc} + N_{mr} + N_{hr}+ N_{hreq}+N_{hreqa} +N_{hc})+\alpha(N_{rcr}+N_{st}+N_{st'} +N_{rc}+ N_{hn}+ N_{crel}+ N_{crel'})+ 2P_{gd}+2P_{ge}+4(P_{gd'}+P_{ge})+5(P_{gd}+P_{ge'}) $
	       $=13m\alpha + 22\beta.$
\item Handover signaling cost for the proposed architecture \\
	       $T_{Handover'}= 2\alpha(N_{mc}+N_{mr}+N_{rcr}+N_{rc})+\alpha(N_{cf'}+N_{mf'}+N_{st}+N_{st'})+ 2P_{ee} + 2P_{ed} + 3(P_{e'e}+ P_{dd})+ P_{e'd}+P_{de}$
	       $=12m\alpha + 12\beta.$
\end{enumerate}
\par We observe that the registration and handover times are lower for the proposed architecture in comparison with the standard 5G cellular architecture mainly due to the reduction in processing cost for encoding and decoding of packet headers. This would depict further improvement if the processing is moved to the core datacenter instead of the less powerful gNBs that are present in the field. The above observation can be quantified by using the values for $\alpha$ and $\beta$ for LTE due to its similarity with the 5G cellular network and the non availability of the values for the 5G cellular wireless system. Using \cite{feasibilityspec}, we have tabulated the calculated values of the KPIs in Table~\ref{table:5gmeas}.

\begin{table}
 \caption{Evaluated reduction in signaling time.}
 \centering
\begin{tabular}{|p{2.1cm}|p{1.6cm}|p{1.7cm}|p{1.6cm}|}
  \hline
  \textbf{System KPI}  & \textbf{\mbox{3GPP 5G} Architecture} & \textbf{\mbox{Proposed 5G} Architecture} &\textbf{Improvement}\\ 
  \hline
  Registration Time & $74-84$ms & $60$ms & $12\%-28\%$ \\
  Handover Time & $78.5$ms & $55.5$ms & $29.29\%$ \\
  \hline 
\end{tabular} 
\label{table:5gmeas}
\end{table} 

\section{Simulations and Results}
\par
\begin{figure}
\centering
\includegraphics[width=0.28\textwidth]{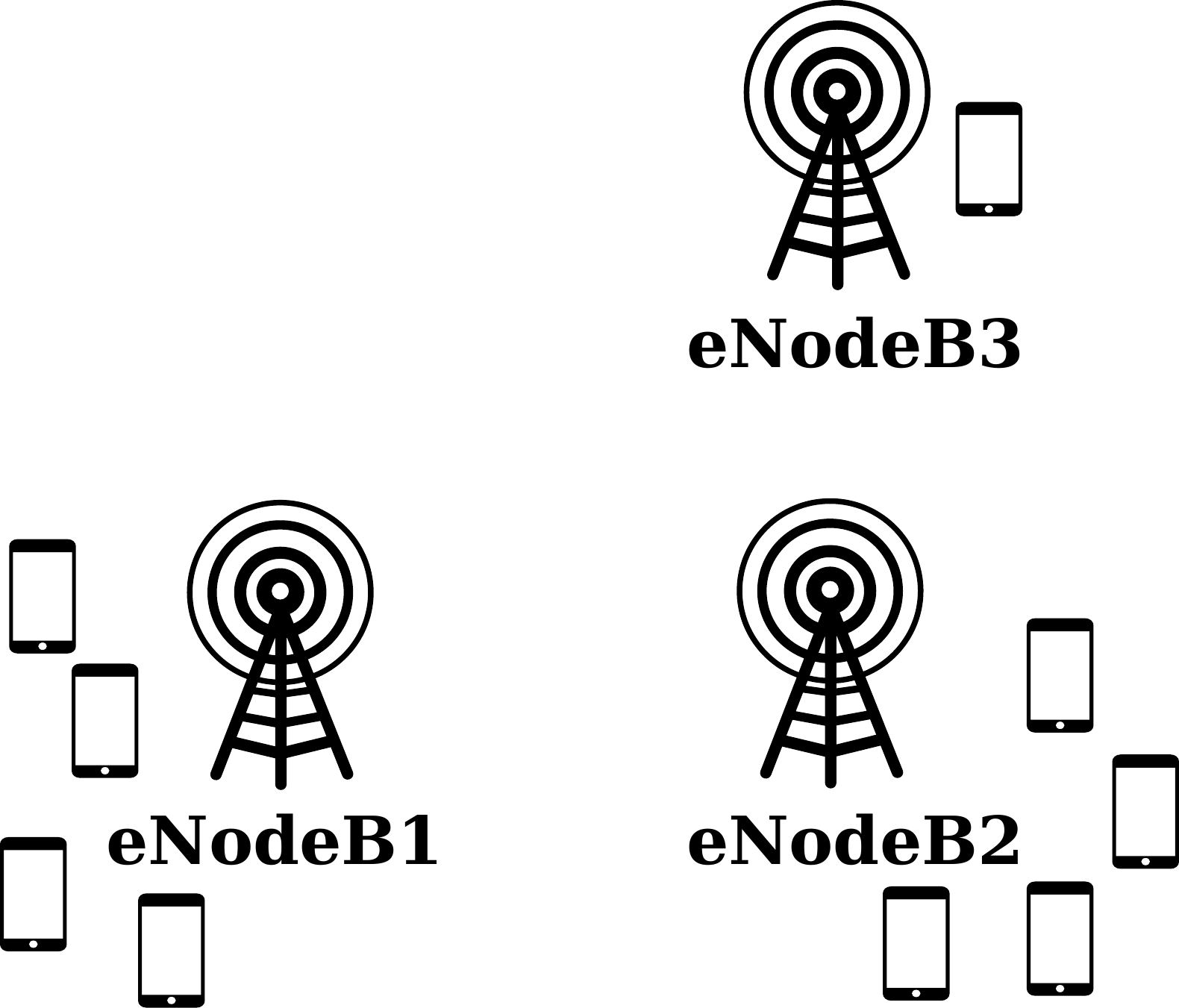}
\caption{Example deployment Scenario.}
\label{fig:handoverscenario}
\end{figure}

\begin{figure}
\centering
\includegraphics[width=0.45\textwidth]{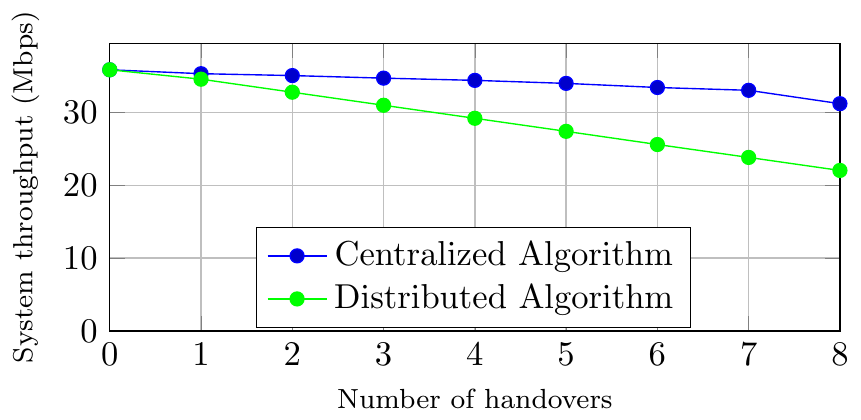}
\caption{System throughput comparison for centralized SDN versus traditional distributed LTE architectures.}
\label{fig:thruput}
\end{figure}

 In order to evaluate the comparative performance of both the architectures, we have carried out simulations with the help of the ns-3 LENA module~\cite{ns3}, as there are no tools available for 5G architecture simulation at present. We have validated the signaling cost improvement for SDN vis-a-vis traditional LTE  by measuring comparative times taken for the attach procedure (in place of Registration). We have also quantified the improvement in the system throughput due to the use of a centralized algorithm for mobility management in place of traditional distributed algorithms.\\ 
\textit{Attach time evaluation: }We have measured the attach time for a single UE using the realtime simulation mode of ns-3. We have observed $3.23$ms and $2.94$ms as the average attach times for the 3GPP defined and the proposed network, respectively. From these estimates, we can observe that the signaling time is reduced by $10\%$. Note that the simulator implements the S1-C interface as an abstraction. Moreover, delays due to the air interface processing are also not taken into account. As a result, the measured times are scaled down in comparison with the real world estimates but the relative performance gain remains the same. As described in the previous section, we can infer that attach time has been reduced due to the reduction in processing time used for encoding and decoding.\\
 \textit{Mobility Management: }Consider a scenario with three Macro eNodeBs, each having a bandwidth of $5$Mhz and transmitting at $46$dBm, placed in the vicinity of each other. As shown in the Figure~\ref{fig:handoverscenario}, eNodeB1 and eNodeB2 are closer to each other with a distance of $400$m and are heavily loaded. eNodeB3 is $500$m away from eNodeB1 and is lightly loaded. We consider a Lognormal pathloss model in the simulation. Consider a vehicular user with a $2$Mbps connection, moving away from eNodeB1 towards eNodeB2 with a speed of $20$m/s. In the traditional X2-based A3 Reference Signal Received Power (RSRP) algorithm, the user is handed over to eNodeB2 as the user received signal strength from the eNodeB2 is the highest. This algorithm runs in a distributed fashion and does not possess load information for all the eNodeBs in the network. As a result, as more and more users move away from the coverage of eNodeB1, they are still handed over to eNodeB2 and the overall system throughput starts deteriorating. In the centralized algorithm, which can be used in the case of SDN aware architectures, the load information along with the RSRP can be used to manage mobility. When experiencing the similar RSRPs from one or more eNodeBs, the users can be handed over to the eNodeB with the lightest load. As a result, the overall system throughput is improved in the face of mobility. As illustrated in Figure~\ref{fig:thruput}, the throughput improvement increases monotonically with the rise in the number of handovers. Hence, we can infer that centralized SDN algorithms perform significantly better in comparison with traditional distributed algorithms in a dynamic environment. 
 
\section{Conclusions}
\par In this paper, we have proposed an SDN based modified architecture for the 5G cellular network in order to centralize the control functionality, and place it in the core network. The movement of RRC functionality together with RRM into the core network, reduces the signaling cost between the NG-RAN and the 5GC. It also centralizes the control of radio resources which results in better decision making at the eAMF due to the network-wide view. The elimination of the NG-AP layer due to the displacement of RRC protocol from the gNB results in the reduction of processing time required for encoding and decoding of header data. We have evaluated the improvement in latency for control plane procedures i.e., registration and handover through performance analysis for both the procedures and simulations for attach time. We have also demonstrated that centralization of the RRC layer and RRM functions leads to better system throughput due to improved mobility management in a dynamic environment.

\bibliography{5Gforum}

\end{document}